# Enhanced User Authentication through Trajectory Clustering

Hazarath Munaga[*,1], J. V. R. Murthy[1], and N. B. Venkateswarlu[2]
[1]Dept. of CSE, JNTU Kakinada, India
Email: {hazarath.munaga, mjonnalagedda}@gmail.com
[2] Dept. of CSE, AITAM, Tekkali, India
Email: venkat_ritch@yahoo.com

*Abstract*— Password authentication is the most commonly used technique to authenticate the user validity. However, due to its simplicity, it is vulnerable to pseudo attacks. It can be enhanced using various biometric techniques such as thumb impression, finger movement, eye movement etc. In this paper, we concentrate on the most economic technique, based on the user habitual rhythm pattern i.e. *not what they type but how they type* is the measure for authenticating the user. We consider the latency between key events as the trajectory, and trajectory clustering is used to obtain the hidden patterns of the user. Obtained pattern can be considered as a cluster of measurements that can be used to differentiate from other users. We evaluated the proposed technique on the data obtained from the 100 users.

## I. INTRODUCTION

As a computer engineer, the main goal is to protect the information or resources from unauthorized users. In literature there are several methods to authenticate validity of the user varying from the usage of biometrics to smart cards. Among them, password authentication is the most acceptable and widely used mechanism because of its economical, operational and implementation advantages. The method relies on the fact that only the authorized user knows the correct password. There is no security in the use of passwords if an impostor knows the password. Hence, to improve the security of user authentication one option is to replace the passwords with a biometrics identification of the user. Currently, there are three major forms of biometrics: physiological, behavioral, and token based. Physiological based biometrics rely on biological attributes such as fingerprints, iris, and retina patterns. Behavioral biometrics utilize behavioral attributes such as voice, signature, and keystroke dynamics. Token-based systems require the possession of a security device such as an ID card. Behavioral biometrics works on the way we interact with the authentication system.

The most popular systems are based on voice, signature, and keystroke dynamics, out of these, keystroke dynamics is purely software-based, it is less expensive and more user transparent. As Gaines et al. [1] observation, a user's keystroke pattern is highly repeatable and distinct from that of other users (typing biometric), which can be used to discriminate the owner from impostors. Hence, typing biometrics based authentication uses an individual's unique typing pattern to validate the authentic user among impostors. The action of typing the password can be analyzed with respect to its physiological characteristics i.e. the latency time between keystrokes, keystroke pressure, key displacement, and key displacement duration.

Ref. [1] introduced the use of keystroke timings as a means of authentication using seven professional typists. Since then there have been a number of research studies [2-9] on authentication of users based on keystroke timings using various techniques ranging from deterministic algorithms to machine learning, and clustering algorithms as learning algorithms for classification.

Ref. [2] develops a profile using statistical analysis method involving means and standard deviations of latencies between consecutive keystrokes. Under these types of statistical models, if a user were to type each key much faster than usual, then he would most likely be rejected because the timing measurement of each of his pairs of consecutive keystrokes would fall beyond the stored mean of his trained profile.

In 1997 Monrose and Rubin use the Euclidean Distance and probabilistic calculations based on the assumption that the latency times for one-digraph exhibits a Normal Distribution [3]. Afterwards, in 2000, they also present an algorithm for identification, based on the similarity models of Bayes, and in 2001 they present an algorithm that uses polynomials and vector spaces to generate complex passwords from a simple one, using the keystroke pattern [4].

In 2000 [5] demonstrated using neural network (NN) novelty detection model, which was built by training the owner's patterns only, and the model was used to detect impostors using some sort of similarity measure, reporting a 1.0% false rejection rate (hereinafter, FRR) and 0% false acceptance rate (hereinafter, FAR). Sung et al [6] has also applied NN to keystroke dynamics, generating error rates on the order of 2-4%. However, such solution suffers from typical NN limitations, e.g.,

---

[*] Hazarath Munaga *alias* MHM Krishna Prasad

conditional independence (i.e., being in a state depends only on the previous state).

Revett et al. have used the rough sets induction algorithm to extract rules that form models for predicting the validity of a login ID/password attempt [7]. The results indicate that the error rate can be as low as 2% in many cases. In addition, the use of a multiple sequence alignment algorithm has been successfully deployed to authenticate a group of users with virtually 100% success [8].

Ref. [9] demonstrated the usage of k-means clustering for validating the user using key stroke dynamics. Whereas, the k-means (i) is not suitable for generating non-globular clusters and, (ii) it is difficult to compare the quality of the computed clusters (e.g. the different initial partitions and the k value affect the outcome). Finally, the k-means approach requires lot of computational time for convergence.

The use of key stroke dynamics based user authentication is no longer a novel concept, but, novelty in this study is the adopted dissimilarity measure and the technique used to identify the user from impostors.

Ref. [11] demonstrated the usage of trajectory clustering for visualizing, analyzing and obtaining hidden patterns from user navigations obtained from virtual environments. Ref. [12] demonstrated the usage of trajectory clustering for selecting cluster heads which implicitly used to extenuate the life time of wireless sensor networks. In this study, we employ a trajectory based clustering algorithm for authenticate the legitimate user based on the two explicit (*key pressed, key released*) and one implicit (*key typed*, which is used to obtain the applied pressure on the key) events.

## II. TRAJECTORY CLUSTERING ALGORITHM

The success of any clustering algorithm depends on the adopted dissimilarity measure. This section explains about the adopted dissimilarity measure.

Ref. [13], proposed the usage of Euclidean distance between time series of equal length as the measure of their similarity. The idea has been generalized in [14] for subsequence matching. In a similar way [15] used Discrete Wavelet Transform and [16] used Principal Component Analysis for measuring time series similarity.

Another approach which is brought from image processing is time warping technique and it is used in [17] to match signals in speech recognition. Berndt and Clifford [18] suggested this technique to measure the similarity of time-series data in data mining. Recent works have also used this similarity measure [19][20].

Ref. [9] suggested the usage of Canberra distance for finding the distance between user samples. Here we used Hausdorff measure [21] for calculating dissimilarity between trajectories, and observe that the Hausdorff measure is more sensitive for small changes than the Canberra distance, which we can observe clearly from the Fig. 1 (obviously, trajectories A and B are overlapped due to its similar nature). Note that, we consider the key stroke latency id also for calculating the Hausdorff dissimilarity.

| Label | Trajectory |
|---|---|
| A | {206,232,192,212,210,168,277} |
| B | {206,232,192,212,210,168,277} |
| C | {216,242,202,222,220,178,287} |
| D | {254,285,135,120,190,228,350} |
| E | {190,220,160,175,235,248,312} |

Fig. 1(a) Sample Trajectories

| Trajectory | A | B | C | D | E |
|---|---|---|---|---|---|
| A | 0 | 0 | 0.16 | 0.03 | 0.05 |
| B | 0 | 0 | 0.16 | 0.03 | 0.05 |
| C | 0.16 | 0.16 | 0 | 0.19 | 0.11 |
| D | 0.03 | 0.03 | 0.19 | 0 | 0.09 |
| E | 0.05 | 0.05 | 0.11 | 0.09 | 0 |

Fig. 1(b) Canberra dissimilarity for Sample Trajectories

| Trajectory | A | B | C | D | E |
|---|---|---|---|---|---|
| A | 0 | 0 | 8.22 | 27.73 | 11.8 |
| B | 0 | 0 | 8.22 | 27.73 | 11.8 |
| C | 8.22 | 8.22 | 0 | 28.56 | 10.95 |
| D | 27.73 | 27.73 | 28.56 | 0 | 21.34 |
| E | 11.8 | 11.8 | 10.95 | 21.34 | 0 |

Fig. 1 (c) Hausdorff dissimilarity for Sample Trajectories

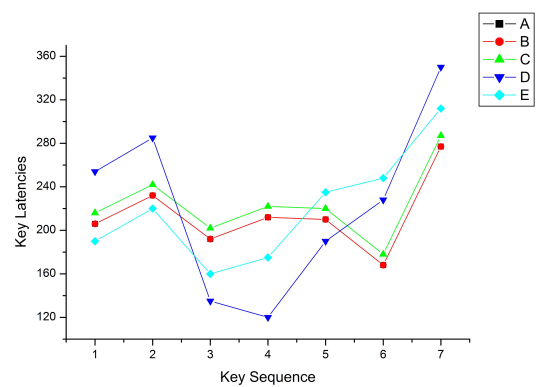

Fig. 1(d) Visualization of the sample trajectories

The following are the some definitions used in our algorithm.

**Definition 1** A *trajectory (t)* is represented as *trj($t_{id}$, $u_0$, $u_1$, $u_2$ .. $u_n$)*, where $t_{id}$ is a unique trajectory id (user id), and $<u_0, u_1, u_2 \ldots u_n>$ is a sequence of key events reflecting the key stroke latencies.

**Definition 2** We define the *key stroke dissimilarity* function between two trajectories $t_1$ and $t_2$ as the maximum of one way distances between two trajectories.

As mentioned in [12], the one way distance from a trajectory $t_1$ to another trajectory $t_2$ is defined as the integral of the Hausdorff distance between points of $t_1$ to trajectory $t_2$ divided by the number of points ($|t_1|$) in $t_1$.

$$dist_{ow}(t_1,t_2) = \frac{1}{|t_1|} \int_{p \in t_1} d_h(p,t_2)dp$$

The Hausdorff distance from a trajectory point $p$ to another trajectory $t_2$ is defined as
$$d(p,t_2) = \min_{q \in t_2}\{d(p,q)\}$$

The distance between trajectories $t_1$ and $t_2$ is the maximum of their one way distances.

$dist(t_1,t_2) = maximum\{dist_{ow}(t_1,t_2), dist_{ow}(t_2,t_1)\}$

Clearly the $dist_{ow}(t_1,t_2)$ is not symmetric but $dist(t_1,t_2)$ is symmetric. Note that $dist_{ow}(t_1,t_2)$ is the integral of the shortest distances from points in $t_1$ to $t_2$.

## A. Generalized Trajectory Cluster Routine

Trajectories are grouped into clusters using the *threshold*. Here the threshold is considered as a maximum value, such that all trajectories (belongs to each user) are grouped into a single cluster. The trajectory cluster routine contains the following stages:

1. Dissimilarity matrix for trajectories will be computed using the Hausdorff distance,
2. Using *Initialization* Algorithm (Table. 1) trajectories are grouped into initial clusters;
3. Using *RepTraj* Algorithm (Table. 1) representative trajectories are computed.
4. By considering the trajectories received from step 3, as initial cluster centres, using *Re-cluster Algorithm* (Table. 1) re compute clusters and their representative trajectories until there is no change in the representative trajectories.

TABLE I
TRAJECTORY CLUSTERING ALGORITHM

| |
|---|
| *Algorithm – 1 (Initialization)* <br> a. Take first sample as first cluster. Classify all the remaining trajectories into this cluster if they are within the threshold. <br> b. Take a trajectory (sequentially) which is not already classified into any of the cluster and consider it as a new cluster. Take all the other trajectories which are not kept in any of the clusters and keep in this cluster if they satisfy the threshold limit. <br> c. Repeat step b till no new clusters are added. |
| *Algorithm – 2 (Representative Trajectory of Cluster C)* <br> For each Trajectory of cluster C calculate cumulative dissimilarity with all other trajectories of the same cluster C. Select the trajectory which is having minimum cumulative dissimilarity and take this as representative trajectory of that cluster. |
| *Algorithm – 3 (Re-compute)* <br> 1. For each Trajectory calculate dissimilarity with all the K representative trajectories and classify to the cluster for which dissimilarity is low (if it is within the threshold). <br> 2. Re-calculate representative trajectories using Algorithm – 2. |

## III. EXPERIMENTAL WORK

Proposed algorithm is experimented on the dataset obtained from 100 computer science undergraduate and graduate students of the JNTUK[2]. The students' are asked to enroll and authenticate during one month period of this study. The users are asked to enter there registration number as the login ID and their self chosen random length string as the password. From the obtained samples, for each user, a user feature is created in the following way:

1. Each participant is asked to type his/her password five times in three random sessions.
2. From each session, using the above trajectory clustering algorithm, a *repTraj* is generated, like this, three *repTraj* are generated from the three random sessions.
3. Average dissimilarity between the three *repTraj* is considered as the *userThreshold*, and by clustering the three *repTraj* (obtained from step 2) a *userRepTraj* is generated.
4. The *userRepTraj* and the *userThreshold* are considered and stored as the user *feature*.

After obtaining the user feature, the participants are asked to face the authentication phase for computing FRR and FAR. For authenticating the user, as usual, the user is asked to enter his login ID and password, from the entered data, a trajectory is generated and dissimilarity is computed with the *userRepTraj*, if the dissimilarity is within the *usersThreshold*, then the user is considered as the authorised user.

Following Fig. 2, shows the obtained visualizations from the tool. Fig. 2-(i) shows the normal (familiar user with the computer) user visualization, in Fig. 2-(i) a, b, c shows the three random sessions and the red highlighted trajectory is the *repTraj* of the session respectively, and highlighted one in Fig. 2-(i)d is the *userRepTraj*, which is obtained by clustering the three *repTraj* shown in *(i)-a, b* and *c*. Similarly, Fig. 2-(ii) shows the example case of a novice user (with the computer) where the FRR is observed.

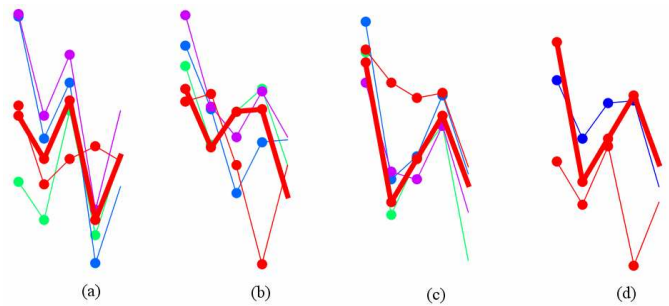

Fig 2. (i) Visualizations of the Normal user

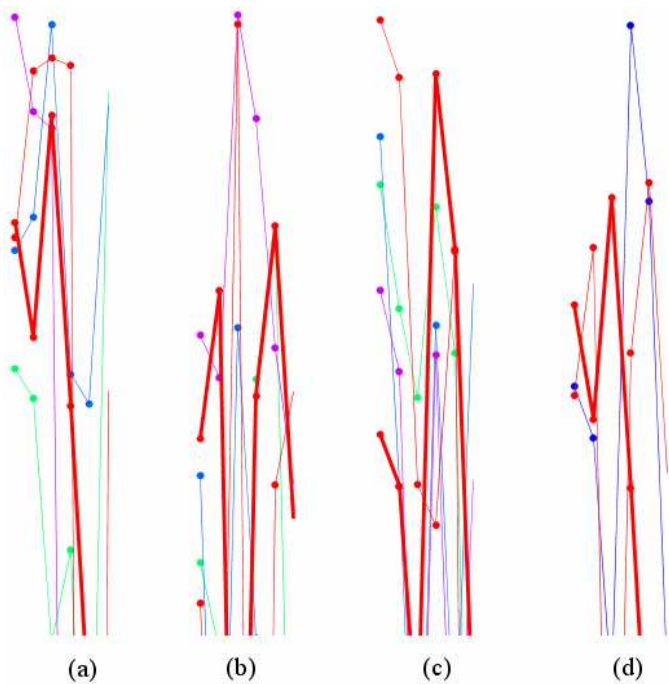

Fig. 2 (ii) Visualizations of novice user, where the FRR rate is observed

---

[2] http://jntukakinada.edu.in;
dataset is available: http://sites.google.com/site/munaga71.

## IV. CONCLUSIONS

In this paper, we have presented a novel trajectory clustering technique for authenticating the user using keystroke dynamics. We have demonstrated the effectiveness of our solution by testing the proposed technique on the data obtained from the 100 computer science undergraduate and graduate students. As discussed in section 3, the technique proved to be efficient in supporting the administrator for authenticating the valid user among other users, even though the password theft occurred. Moreover, for the typing biometrics system, it was observed that the Hausdorff dissimilarity measure can be adopted for getting effective results, *i.e.* as shown in section 2 the Hausdorff dissimilarity measure is more sensitive for small variations. As a limitation of the proposed technique, during experimentation phase, we tested the toll on a novice user, where we observed the False Rejection Rate.

Even though the continuous monitoring concept of key stroke dynamics is available, but it can not be applicable for all applications (e.g. accessing ATM, where only small amount of time/keys used to access the system), this static concept of key stroke dynamics will be useful. User authentication supported by Keystroke dynamics has many applications in the today's electronic world especially applications where secure data transfer is mandatory.